\documentstyle[sprocl]{article}

\newcommand{\eg}{{e.g.}}

\newcommand{\CP}{$C\!P$}
\newcommand{\etal}{{\em et al.}}

\begin{document}

\title{
\begin{flushright}
{\normalsize
IIT-HEP-98/3\\
September 1998\\
\vspace{0.1in}}
\end{flushright}
EXPERIMENTAL PROSPECTS FOR \CP\ VIOLATION IN CHARM\footnote
{Invited talk presented at the {\em Workshop on CP Violation},
3--8 July 1998, Adelaide, Australia}
}

\author{DANIEL M. KAPLAN}
\address{Physics, Illinois Institute of Technology,\\
Chicago, IL 60616, USA\\
E-mail: kaplan@fnal.gov}

\author{VASSILI PAPAVASSILIOU}
\address{Physics, New Mexico State University,\\
Las Cruces, NM 88003, USA\\
E-mail: pvs@nmsu.edu}

\author{for the BTeV Collaboration}
\address{ }


\maketitle

\begin{abstract} 
Experimental sensitivity to \CP\ violation in charm decay is beginning to
approach the interesting regime ($\sim10^{-3}$) in which new physics may be
manifest. In the early years of the 21st century,
if the technical challenges can be met, 
the proposed BTeV experiment should
have the best sensitivity for rare effects both in
charm and in beauty.
\end{abstract}

\section{Charm \CP\ Violation in the Standard Model and Beyond}

\begin{figure}
\label{fig:penguin}
\vspace{-0.15in}
\hspace{-0.08in}\centerline{\epsfxsize 4.85 in\epsffile{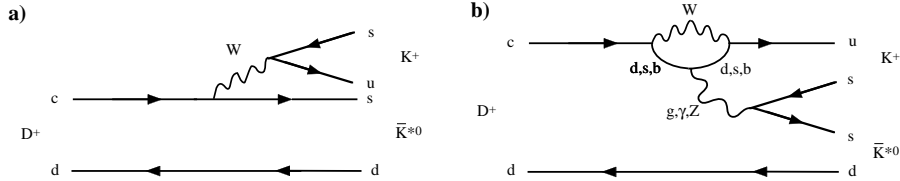}}
\caption{Example of Cabibbo-suppressed $D^+$ decay that can proceed via  both
(a) tree and (b) penguin diagrams.}
\end{figure}

The Standard Model (SM) predicts direct \CP\ violation at the ${\cal
O}(10^{-3})$ level in singly Cabibbo-suppressed (SCS) charm
decays,\cite{charmCP} due to the interference of tree-level processes with
penguins (Fig.~\ref{fig:penguin}). The observation of \CP\ asymmetries
substantially larger than this could be unambiguous evidence of new physics, as
would almost any\footnote{In $D^0$ (but not charged-$D$) 
decays, ${\cal O}(10^{-3})$
\CP\ asymmetries may be possible in the SM due to interference 
between DCS and mixing amplitudes.}
observation of \CP\ violation in Cabibbo-favored (CF) or doubly
Cabibbo-suppressed (DCS) charm decays.

A variety of extensions of the Standard Model have been considered\,\cite{Nir} 
in which 
charm \CP\ asymmetries could be as large as ${\cal O}(10^{-2})$. These include
models with leptoquarks,\cite{Lepto} extra Higgs doublets (\eg\ non-minimal
supersymmetry\,\cite{Bigi94}), a fourth generation,\cite{Pakvasa,Babu} or 
right-handed weak currents.\cite{Pakvasa,Yaouanc} In addition, two Standard
Model
possibilities for large \CP\ asymmetries in charm have been discussed: 
asymmetries due to $K^0$ mixing in \eg\ $D^\pm\to K_S\pi^\pm$,\cite{Xing} and 
the possibility that $D$ mesons mix with glueballs or gluonic 
hybrids.\cite{Close-Lipkin}

\section{Limits on Charm \CP\ Violation}

Exponentially-increasing charm event samples have led to substantially 
improved \CP-violation limits over time. The most sensitive limits 
come from Fermilab fixed-target experiments E791\,\cite{Aitala} and 
E687\,\cite{Frabetti} and from CLEO II.\cite{Bartelt} These have been combined 
into world averages by the Particle Data Group for the 1998 Review of Particle 
Physics (Table~\ref{tab:limits}).\cite{PDG98} No significant signals have 
been observed, and most limits are in the range of several percent.
There is thus a substantial discovery window for new physics in SCS modes, and 
an even larger one for CF and DCS modes (for which almost no limits are 
available\,\cite{Bartelt-comment}).

\begin{table}
\caption{World-average charm \CP\ asymmetries 
(from Ref.~\protect\cite{PDG98}).}
\vspace{3mm}
\label{tab:limits}
\begin{center}
\begin{tabular}{|ccc|}
\hline
Particle & Mode & Asymmetry \\ \hline
$D^\pm$ & $K^+K^-\pi^\pm$ & $-0.017\pm0.027$ \\
 & $K^\pm K^{*0}$ & $-0.02\pm0.05$ \\
 & $\phi\pi^\pm$ & $-0.014\pm0.033$ \\
 & $\pi^+\pi^-\pi^\pm$& $-0.02\pm0.04$ \\ \hline
${}^{{}^(}\overline{D}{}^{{}^)}\!{}^0$ & $K^+K^-$ & $0.026\pm0.035$ \\
 & $\pi^+\pi^-$ & $-0.05\pm0.08$ \\
 & $K_S\phi$ & $-0.03\pm0.09$ \\
 & $K_S\pi^0$ & $-0.018\pm0.030$ \\ \hline
\end{tabular}
\end{center}
\end{table}

These time-integrated \CP\ asymmetries are defined as
\begin{equation}
A_{CP}=\frac{\Gamma(D\to f) - \Gamma({\overline D}\to\overline{f})}
{\Gamma(D\to f) + \Gamma({\overline D}\to\overline{f})}\,.
\end{equation}
Note that while not explicitly time-dependent, $A_{CP}$ is sensitive to decay 
time through the vertex-separation cuts used to suppress non-charm background. 
For $D^0$ decays, it is thus sensitive to both direct and indirect 
\CP\ violation.\cite{Xing2}

When (as in fixed-target experiments) the initial state is non-\CP-symmetric, 
the observed rate asymmetry in a given mode must be corrected for 
$D$-${\overline D}$ production asymmetry. E687 and E791 therefore normalize 
their observed rates in SCS modes to those in CF modes, for example, 
\begin{equation}
A_{CP}\left({}^{{}^(}\overline{D}{}^{{}^)}\!{}^0\right)=
\frac{
\frac{N(D^0\to f)}{N(D^0\to K^-\pi^+)}-\frac{N(\overline{D}{}^0\to 
\overline{f})}{N(\overline{D}{}^0\to K^+\pi^-)}
}
{
\frac{N(D^0\to f)}{N(D^0\to K^-\pi^+)}+\frac{N(\overline{D}{}^0\to 
\overline{f})}{N(\overline{D}{}^0\to K^+\pi^-)}
}. \label{eq:ratio}
\end{equation}
Acceptances and efficiencies tend to cancel in 
Eq.~\ref{eq:ratio}, reducing systematic uncertainties.

\section{Prospects for Improved Sensitivity}

Several experiments soon to take data are expected to surpass 
current sensitivities. Current and future experiments are summarized in
Table~\ref{tab:exp}. Since \CP-violation sensitivity
depends in complicated ways on reconstruction and particle-ID
efficiency for various modes, optimization of vertex cuts, 
$D^*$-tagging efficiency (for $D^0$ modes),
etc., we use here simple overall benchmarks rather
than detailed estimates. These are total number of charm decays
produced or reconstructed and total number of
${}^{{}^(}\overline{D}{}^{{}^)}\!{}^0\to K^\mp\pi^\pm$ decays produced or
reconstructed. We scale from current experiments according to the square root
of one of these benchmark numbers to obtain an estimated \CP\ reach in each
case, recognizing that this procedure is at best approximate and addresses
only the statistical component of \CP\ sensitivity.

\setcounter{footnote}{0}
\renewcommand{\thefootnote}{\fnsymbol{footnote}} 

\begin{table}
\caption{Current and approved future experiments with charm \CP-violation 
sensitivity.}
\label{tab:exp}
\vspace{3mm}
\begin{center}
\footnotesize
\begin{tabular}{|l|cc|cc|c|}
\hline
& \multicolumn{2}{|c|}{All charm decays} & 
\multicolumn{2}{|c|}{${}^{{}^(}\overline{D}{}^{{}^)}\!{}^0\to K^\mp\pi^\pm$} & 
$\sigma(A_{CP})$ \\
\raisebox{1.5ex}[0pt]{Exp't} & prod. & rec. & prod. & rec. & (SCS) \\ \hline
FNAL E687 & & $0.8\times10^5$ & & & $\approx0.1$ \\
FNAL E791 & $10^8$ & $2.5\times10^5$ & $1.2\times10^6$  & $3.7\times10^4$ & 
$\approx0.05$ \\
CLEO II\footnotemark & $2.7\times10^6$  & & $1.0\times10^5$ & $1.8\times10^4$ & 
$\approx0.05$\\ \hline
FOCUS (FNAL E831) & & $10^6$ & & & $\approx0.03$? \\
COMPASS & & & & $7\times10^4$ & $\approx0.03$? \\
HERA-$B$ & few$\,\times10^{10}$/yr & & & & ? \\
$B$ Factories, CLEO III & $3\times10^7$/yr & & & & $\approx0.01$? \\ \hline
\end{tabular}
\end{center}

\vspace{3mm}
\baselineskip=8 pt
\setcounter{footnote}{0}

{\footnotesize
\footnotemark
CLEO II sensitivity given as of Ref.~\cite{Bartelt}; additional data are being 
accumulated, and the final CLEO II sample should be 
substantially larger.
}
\end{table}

The $B$ Factories and CLEO III are expected to have the best charm \CP\ reach
among approved future experiments. (While HERA-$B$ has the highest charm
production rate, it is likely to have poor trigger efficiency for charm due to
$p_t$ requirements imposed at the trigger level.) 
In multi-year runs, the combined reach of
these experiments could be approximately an order of magnitude better
than current limits, reaching the few$\,\times10^{-3}$ level.

We have seen above that this sensitivity is unlikely to be sufficient to
observe Standard Model \CP\ violation in charm, though it may suffice for the
discovery of non-SM effects if they are large. As we will see, the proposed
BTeV experiment at the Tevatron should be able to achieve 
$\sim10^{-4}$ sensitivity, bringing even SM effects within reach.

\section{The BTeV Experiment}

BTeV is an approved R\&D program at Fermilab aimed at proposing a
collider charm and beauty experiment for
Tevatron Run II and beyond. Its main physics goals
are to search for \CP\ violation, mixing, and rare flavor-changing
neutral-current decays of beauty and charm at unprecedented
levels of sensitivity. Each year of BTeV collider operation is expected to
produce ${\cal O}(10^{11}$) $b$ hadrons and ${\cal O}(10^{12}$) $c$ hadrons, to
be compared with ${\cal O}(10^{7}$) of each available at the $B$ Factories and
${\cal O}(10^{9}$) and ${\cal O}(10^{10}$) per year at HERA-$B$. The BTeV
spectrometer is being designed to make optimal use of the produced samples,
avoiding many of the compromises necessary in general-purpose detectors.

Since $B$ physics is a major goal of BTeV, we here summarize projected
sensitivities for beauty as well as charm physics.
More
detailed discussions, both of the  proposed apparatus and of its physics reach,
may be found in Refs.~\cite{eoi,ICHEP98}.

\subsection{The BTeV Spectrometer}

\begin{figure}
\vspace{-.5in}
\centerline{
\epsfxsize=4.5 in\epsffile{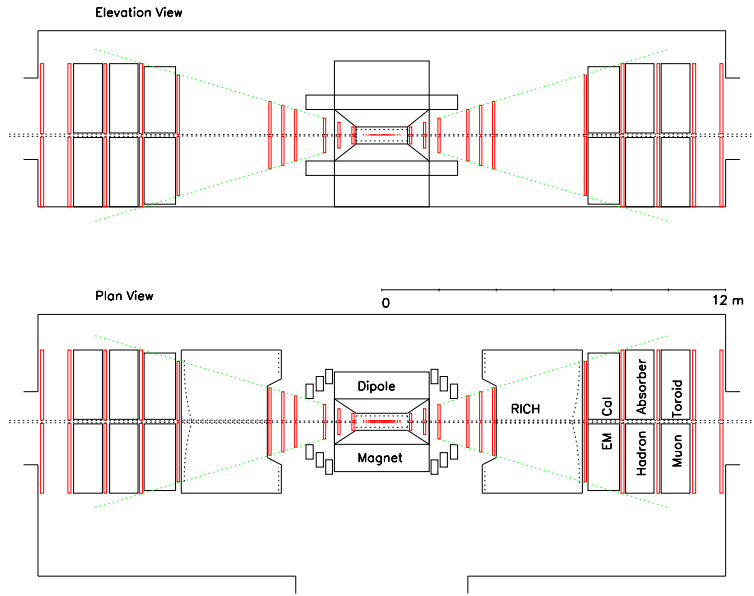}}
\vspace{-1.1 in}
\caption{Elevation and plan of the BTeV spectrometer.\label{fig:spect}}
\end{figure}

\begin{figure}
\vspace{-0.4 in}
\centerline{\epsfxsize=3.25 in \epsffile{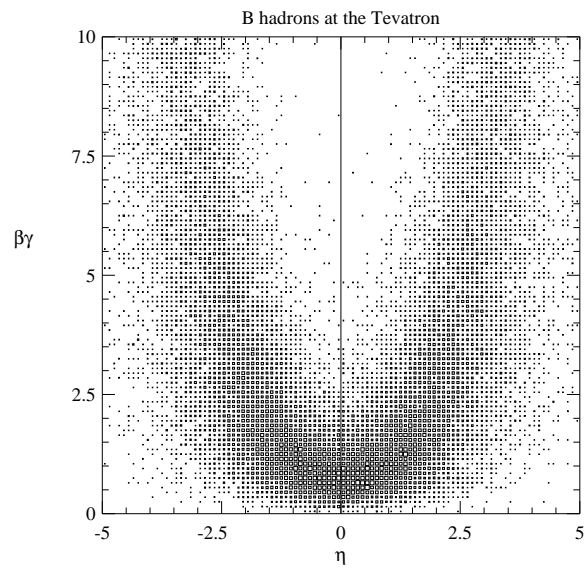}}
\vspace{-.1in}
\caption{Relativistic boost factor $\beta\gamma$ {\em vs.}\ pseudorapidity 
$\eta$ of $B$ hadrons produced at the Tevatron Collider.\label{fig:bg-vs-y}}
\end{figure}

The proposed BTeV spectrometer (Fig.~\ref{fig:spect}) covers the 
forward and backward regions at the new C0 Tevatron interaction area. The 
instrumented angular range is 
0.01\,$_\sim$\llap{$^<$}$\,|\tan{\theta}|\,_\sim$\llap{$^<$}\,0.3, 
corresponding to the approximate pseudorapidity range
1.5\,$_\sim$\llap{$^<$}$\,\eta\,_\sim$\llap{$^<$}\,6 for the parent particle.
Monte Carlo simulation shows that such coverage gives $\approx$10--50\%
acceptance (depending on mode) for $B$ and $D$ decays.
 Compared to the ``central-geometry" case (\eg\ CDF and D0), this 
``forward-geometry" configuration accepts relatively high-momentum particles 
(see Fig.~\ref{fig:bg-vs-y}), allowing 
better reconstruction of decay proper time. Another advantage is
the feasibility of effective charged-hadron identification.

Because QCD
mechanisms of $b\bar b$ production yield quark pairs that are closely
correlated in pseudorapidity ($|\eta_b-\eta_{\bar b}|\,_\sim$\llap{$^<$}\,1),
there is little
disadvantage in omitting the small-$\eta$ region: when the decay products of
one $B$ hadron are detected in the forward (or backward) region, decay products
of the second (``tagging") $B$ have a high probability to be detected there
also. (And of course, for ``same-side" tagging\,\cite{CDF-tagging} the direction
of the other $B$ is immaterial.)

In addition to large acceptance, the apparatus must have high interaction-rate
capability, superb vertex reconstruction,
an efficient trigger, high-speed and  high-capacity data
acquisition, good mass resolution, and good particle 
identification. Of these requirements, the most challenging are the vertexing, 
the trigger,
and the particle identification.
It is these challenges that the BTeV R\&D program is addressing.

\begin{figure}
\centerline{\epsfysize=3.25 in \epsffile{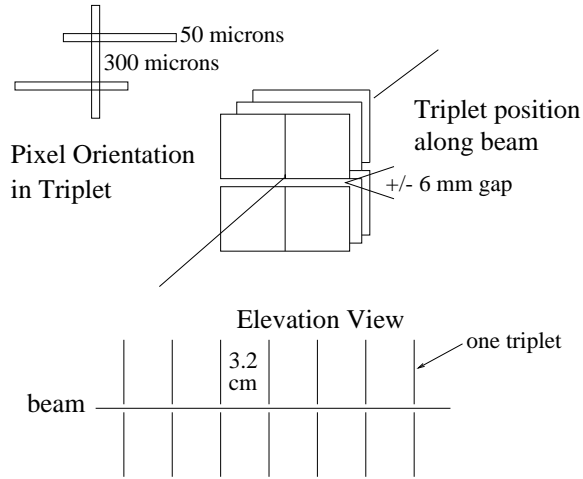}}
\vspace{-.65in}
\caption{Arrangement of vertex detector as proposed in BTeV 
Expression of Interest.\protect\cite{eoi} 
As discussed in the text,
vertex resolution is improved by use of a 12-mm-square beam hole rather than
the horizontal shown here, and the square-hole arrangement has now been
adopted as the BTeV baseline design.\label{fig:vertex}}
\end{figure}

 We intend to trigger primarily on the presence
of a decay vertex separated from the primary vertex.\cite{Vertex-trigger} To
reduce occupancy and facilitate vertex reconstruction at trigger level 1, pixel
detectors (Fig.~\ref{fig:vertex}) will be used for vertex reconstruction. For
efficient, reliable, and compact particle  identification, we will use
ring-imaging Cherenkov counters. In other respects the spectrometer layout will
resemble that of existing large-aperture fixed-target heavy-quark experiments.

A crucial detail of spectrometer design deserves comment, since it has a
large  impact on sensitivity. As the size
of the gap between the upper and lower halves of the vertex detectors is
reduced, for pixel resolution fine enough that multiple
scattering dominates, vertex resolution improves linearly.
However, there is a minimum gap size below which radiation damage to the
pixel  detectors becomes unacceptably large. Given these competing
requirements, we find that resolution is optimized by use of a square beam
hole rather than the horizontal gap shown in Fig.~\ref{fig:vertex}. 

\begin{figure}
\vspace{-.1 in}
\centerline{\epsfxsize=3.25 in 
\vspace{-.1in}
\epsffile{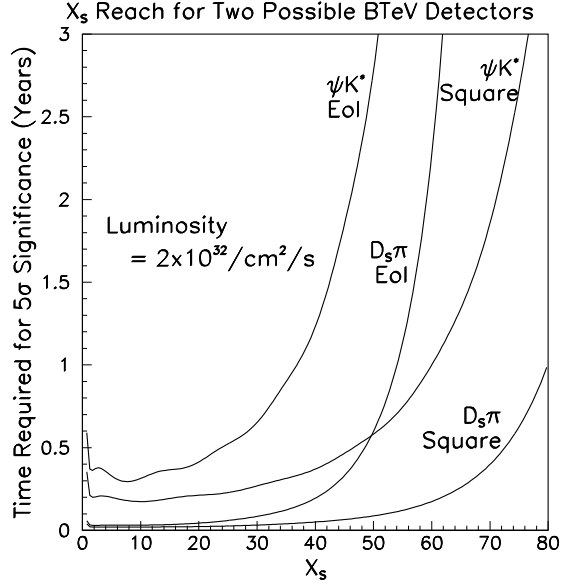}}
\caption{BTeV reach in $x_s$ for two $B_s$ decay modes, comparing square-hole 
configuration with horizontal-gap configuration.\label{fig:mixing-reach}}
\end{figure}

As an example we consider sensitivity to $B_s$ mixing. Improved vertex
resolution helps two ways: both in resolving the extremely rapid
$B_s$-$\overline{B}{}_s$ oscillations, and by enlarging the event sample that
passes the vertex cuts needed to suppress background. This is illustrated in 
Fig.~\ref{fig:mixing-reach}, in which the reach in the $B_s$ mixing parameter 
$x_s$ in the $J/\psi K^*$  and
$D_s\pi$ decay  modes is compared for the ``EoI"\,\cite{eoi} vertex-detector
configuration (with 12-mm  horizontal gap) and for a vertex detector with a
12-mm-square beam hole. The $x_s$ reach is substantially better with the
square hole. For example, if $x_s$ is 60, its determination at $5\sigma$ 
significance would require two months of running with the square hole but 
three years with the horizontal gap.
The square-hole configuration has now been
adopted as the BTeV baseline design.

\subsection{BTeV Beauty Sensitivity}

Especially for nonleptonic final states, BTeV's beauty sensitivity is expected
to surpass that of all other proposed experiments. Since beauty
experiments have many goals, comparing their sensitivities is an involved
procedure.
Table~\ref{tab:bsens} gives a representative set of benchmarks. 

\setcounter{footnote}{0}

\begin{table}
\caption{Representative examples of BTeV $b$-physics reach (from 
Ref.~\protect\cite{ICHEP98}).}
\label{tab:bsens}
\vspace{3mm}
\begin{center}
\begin{tabular}{|cc|}
\hline
Measurement & Accuracy/10$^7$\,s\\
\hline
$x_s$ & $>80$ \\
$\sin{2\beta}$ (using $B^0\to\psi K_S$) & $\pm0.013$ \\
$A_{CP}(B^0\to\pi^+\pi^-)$ & $\pm0.013\footnotemark$ 
\setcounter{footnote}{0} \\
$\gamma$ (using $D_s K^-$) & $\pm\approx8^\circ$ \\
$\gamma$ (using $D^0 K^-$) & $\pm\approx8^\circ$ \\
$B\!R(B^-\to K^-\mu^+\mu^-)$ & $5\times10^{-8}$ (at $4\sigma)\footnotemark$ \\ 
\hline
\end{tabular}
\end{center}

\vspace{3mm}
\baselineskip=8 pt
\setcounter{footnote}{0}

{\footnotesize
\footnotemark
These results are for a vertex detector with a horizontal beam gap, as opposed 
to the square beam hole used in the other simulations.
}
\end{table}

\subsection{BTeV Charm Sensitivity}

BTeV's charm sensitivity depends on running mode.
BTeV can operate both in collider and fixed-target modes. The latter mode is
achieved by suspending a thin wire or small pellet in the halo of the proton
or antiproton beam. Given the accelerator upgrades needed to achieve 
high-luminosity $\overline{p}p$ collisions at C0, fixed-target running may
occur before collider running.  
The huge increase in
$b\bar b$ cross section from $\sqrt{s}=0.043$ to 2\,TeV\,\cite{sigma-b} means
that significant beauty sensitivity is available only in collider mode.
However,  useful charm sensitivity may be available in fixed-target mode. 

\setcounter{footnote}{0}

\begin{table}
\caption{Charm sensitivity in BTeV fixed-target and collider modes.}
\label{tab:FT-coll}
\vspace{3mm}
\begin{center}
\begin{tabular}{|lcc|}
\hline
Quantity & FT & Collider \\ \hline
Running time &	$10^7$ s & $10^7$ s \\
Interaction rate & $2\times 10^6$ s$^{-1}$ & $1.5\times10^7$ s$^{-1}$ \\
${}^{{}^(}\overline{D}{}^{{}^)}\!{}^0$/interaction & 
$6.5\times10^{-4}  A^{0.29}$\footnotemark & 1\%?\footnotemark \\
$A^{0.29}$ &	2 - 4.5 (C - W) & 1 \\
$B\!R(D^0\to K^-\pi^+)$ & 3.85\% & 3.85\% \\
${}^{{}^(}\overline{D}{}^{{}^)}\!{}^0\to K\pi$ produced & 
$(1 - 2.3)\times 10^9$ & $6 \times 10^{10}$? \\
Acceptance & 35\% & 27\% \\
Trigger eff. & 15\% & 11\% \\
Reconst.\ eff.\ & 40\% & 40\% \\ \hline
$D^0\to K\pi$ reconst. & $(2-5)\times10^7$ & $7\times10^8$? \\
\hline
\end{tabular}
\end{center}

\setcounter{footnote}{0}
\vspace{3mm}
\baselineskip=8 pt

{\footnotesize
\footnotemark Extrapolated from measurements at $\sqrt{s}=39\,$GeV.\\
\footnotemark Assumed since no measurement is yet available.\\
}
\end{table}

Table~\ref{tab:FT-coll} compares charm sensitivity in the two running modes.
There is some uncertainty in each case. For example, the optimal choice of 
material for the fixed target is not yet clear, so we consider a range from 
carbon to tungsten. For collider, the charm production cross section has not 
yet been measured, so we use an educated guess. Also, in both cases 
efficiency estimates can be expected to evolve as our simulations become more 
sophisticated. However, the potential is clear.
Scaling from FNAL E791, we may expect \CP\ sensitivity in SCS modes at the
level of a few\,$\times10^{-4}$ per year of collider running. 
If systematic uncertainties can be controlled,
BTeV should be able to observe significant 
\CP\ asymmetries in charm decay even at the Standard Model level.

\section{Conclusions}

It is becoming increasingly clear that full understanding of the mechanisms of
$B$ decay and their bearing on the unitarity triangle will require the large
beauty event samples available only in hadroproduction. Given the complexity
of these analyses,\cite{London} it may be that unexpected effects in charm
decay will provide the first evidence of physics beyond the Standard Model. By
providing large well-measured samples both of beauty and of charm, BTeV could be
the key experiment that will lead to a breakthrough in our understanding in
the early years of the next century.

\section*{Acknowledgements}

DMK thanks the organizers for the opportunity to consider this intriguing 
subject in such congenial surroundings.
This work was supported in part by the U.S. Dept. of Energy and the Special 
Research Centre for the Subatomic Structure of Matter at the University of
Adelaide.


\begin{thebibliography}{99}
\bibitem{Bartelt-comment} 
Although the Particle Data Group does not include it
in their review,\protect\cite{PDG98} Bartelt \etal\protect\cite{Bartelt}
give the first published result on \CP\ violation in the (Cabibbo-favored) 
${}^{{}^(}\overline{D}{}^{{}^)}\!{}^0
\to K^\mp\pi^\pm$
mode: $A_{K\pi}=0.001\pm0.011$.
\end{thebibliography}
\end{document}